\documentclass[aps,prl,twocolumn,footinbib,showpacs,superscriptaddress,nobalancelastpage,floatfix,10pt]{revtex4-2}

\usepackage[english]{babel} 
\usepackage[utf8]{inputenc} 
\usepackage[T1]{fontenc} 
\usepackage{physics} 
\usepackage{siunitx} 
\usepackage{amsmath} 
\usepackage{amssymb} 
\usepackage{amstext} 
\usepackage{graphicx} 
\usepackage[hidelinks]{hyperref} 
\usepackage[dvipsnames]{xcolor} 
\usepackage{lipsum} 
\usepackage{array} 
\usepackage[all]{hypcap} 
\usepackage[normalem]{ulem}
\usepackage[ruled,nofillcomment,linesnumbered]{algorithm2e} 

\AtBeginDocument{\RenewCommandCopy\qty\SI}
\ExplSyntaxOn
\msg_redirect_name:nnn { siunitx } { physics-pkg } { none }
\ExplSyntaxOff

\hypersetup{
    colorlinks=true,
    breaklinks=true,
    urlcolor=black,
    linkcolor=black,
    citecolor=NavyBlue,
    bookmarksopen=true,
    pdftoolbar=false,
    pdfmenubar=false,
}

\usepackage{cleveref}
\crefname{equation}{Eq.}{Eqs.}
\Crefname{equation}{Equation}{Equations}
\crefname{algorithm}{Alg.}{Algs.}
\Crefname{algorithm}{Algorithm}{Algorithms}
\crefname{figure}{Fig.}{Figs.}
\Crefname{figure}{Figure}{Figures}
\crefname{section}{Sec.}{Secs.}
\Crefname{section}{Section}{Sections}
\crefname{appendix}{Appendix}{Apps.}
\Crefname{appendix}{Appendix}{Apps.}
\crefname{paragraph}{Sec.}{Secs.}
\crefname{table}{Table}{Tables}

\newcommand{\lettersection}[1]{\paragraph*{\color{black}#1.---}}

\renewcommand{\a}{\Hat{a}} 
\newcommand{\adag}{\Hat{a}^\dagger} 
\renewcommand{\b}{\Hat{b}} 

\newcommand{\f}{\Hat{f}}
\newcommand{\fdag}{\Hat{f}^\dagger}
\newcommand{\n}{\Hat{n}}
\renewcommand{\H}{\Hat{H}} 
\renewcommand{\L}{\Hat{L}} 
\newcommand{\hrho}{\Hat{\rho}} 
\newcommand{\hphi}{\Hat{\phi}}
\newcommand{\hO}{\Hat{O}}
\newcommand{\Liouv}{\mathcal{L}}
\newcommand{\kket}[1]{\vert#1\rangle\!\rangle}
\newcommand{\bbra}[1]{\langle\!\langle#1\vert}
\newcommand{\bbrakket}[2]{\langle\!\langle#1\vert#2\rangle\!\rangle}

\DeclareMathOperator{\ReLU}{ReLU}

\begin{document}

\title{
Optimal control in large open quantum systems -- \texorpdfstring{\\}{} the case of transmon readout and reset
}

\author{Ronan Gautier}
\affiliation{Institut Quantique and D\'epartement de Physique, Universit\'e de Sherbrooke, Sherbrooke, QC, Canada}
\affiliation{Laboratoire de Physique de l'\'Ecole Normale Sup\'erieure, Inria, ENS, Mines ParisTech, Universit\'e PSL, Sorbonne Universit\'e, Paris, France}
\affiliation{Alice \& Bob, Paris, France}

\author{\'Elie Genois}
\affiliation{Institut Quantique and D\'epartement de Physique, Universit\'e de Sherbrooke, Sherbrooke, QC, Canada}

\author{Alexandre Blais}
\affiliation{Institut Quantique and D\'epartement de Physique, Universit\'e de Sherbrooke, Sherbrooke, QC, Canada}
\affiliation{Canadian Institute for Advanced Research, Toronto, ON, Canada}


\begin{abstract}
We present a framework that combines the adjoint state method together with reverse-time backpropagation to solve prohibitively large open-system quantum control problems. Our approach enables the optimization of arbitrary cost functions with fully general controls applied on large open quantum systems described by a Lindblad master equation. It is scalable, computationally efficient, and has a low memory footprint. We apply this framework to optimize two inherently dissipative operations in superconducting qubits which lag behind in terms of fidelity and duration compared to other unitary operations: the dispersive readout and all-microwave reset of a transmon qubit. Our results show that while standard pulses for dispersive readout are nearly optimal, adding a transmon drive during the protocol can yield 2x improvements in fidelity and duration. We further demonstrate a 2x improvement in reset fidelity and duration through pulse shaping, indicating significant potential for enhancement in reset protocols. Our approach can readily be applied to optimize quantum controls in a vast range of applications such as reservoir engineering, autonomous quantum error correction, and leakage-reduction units.
\end{abstract}

\maketitle

\lettersection{Introduction}

Quantum optimal control (QOC) provides a framework to design external controls for realizing arbitrary quantum operations with maximal fidelity and minimal time~\cite{peirce1988optimal,werschnik2007quantum,koch2022quantum}, crucial requirements of useful quantum error correction~\cite{terhal2015quantum}. A common assumption of QOC is that minimizing operation time will also reduce the impact of environmental noise, such that only closed quantum systems need to be considered. However, these approaches are limited by the fact that controlling the system's coherent dynamics can drastically alter the impact of some noise sources, as exemplified by dynamical decoupling methods~\cite{viola1999dynamical,khodjasteh2005fault,biercuk2009optimized}. Moreover, closed-system approaches cannot extend to inherently dissipative processes such as qubit readout and reset. Consequently, optimally controlling open quantum systems emerges as an important avenue~\cite{schmidt2011optimal,schulte2011optimal,koch2016controlling}. It addresses both the minimization of decoherence in quantum information processing~\cite{goerz2014optimal,an2021quantum,propson2022robust} and the design of dissipative protocols~\cite{egger2014optimal,boutin2017resonator,basilewitsch2019reservoir}, marking a significant step towards comprehensive quantum control in engineered systems.

Over the last decade, several approaches have emerged for open-system QOC. Closed-loop control methods -- such as feedback engineering~\cite{chen2020quantum,porotti2023gradient} and reinforcement learning~\cite{sivak2022model,sivak2022real,baum2021experimental,ding2023high} -- have seen recent success but are difficult to scale to many parameters. Open-loop methods -- including gradient ascent pulse engineering~\cite{khaneja2005optimal,boutin2017resonator}, Krotov's method~\cite{krotov1995global,goerz2019krotov,basilewitsch2019quantum}, and automatic differentiation~\cite{jirari2009optimal,leung2017speedup,baydin2018automatic,abdelhafez2019gradient} -- circumvent this limitation by computing numerical gradients, enabling efficient exploration of the parameter space. However, controlling large open systems remains challenging: the sheer problem size prohibits frameworks that require vectorizing the Liouvillian or storing density matrices at each time step. Although methods for low-memory differentiation have been proposed for closed quantum systems~\cite{goerz2022quantum,lu2024optimal}, extending them to open systems while maintaining a favorable memory overhead remains unresolved.

In this Letter, we present a framework enabling the realization of QOC on large open quantum systems with a fully general parametrization over the controls and arbitrary cost functions. Our approach combines the adjoint state method~\cite{pontryagin1962mathematical,dehaghani2023application,boscain2021introduction} with reverse-time backpropagation~\cite{chen2018neural,kidger2022neural,somloi1993controlled,narayanan2022reducing} to reduce the memory cost of differentiation from linear to constant relative to the number of numerical integration steps. This significant reduction enables solving otherwise prohibitively large open-system quantum control problems defined in Lindblad form~\cite{breuer2002theory}, and makes the scheme ideal for GPU acceleration. Our approach thus ensures precise, fast and scalable computation of arbitrary gradients. We apply this method to optimize two critical operations for the realization of a fault-tolerant quantum computer based on superconducting circuits: dispersive readout~\cite{blais2004cavity,walter2017rapid,sunada2022fast} and all-microwave reset~\cite{pechal2014microwave,magnard2018fast} of a transmon qubit~\cite{koch2007charge}. Using the signal-to-noise ratio (SNR) of the readout as a cost function, we find several optimal controls with increasing experimental complexity and up to 2x improvements in fidelity and duration compared to standard protocols. For reset, we show that pulse shaping alone can halve the operation duration, an improvement of high practical relevance.

\lettersection{Adjoint state method}
Consider a QOC problem for which we seek to find a set of parameters minimizing a cost function $C(\theta, \hrho(t_0), \ldots, \hrho(t_n))$. This function, in general, depends on both the problem parameters $\theta = (\theta_1, \ldots, \theta_m)$ and on the density matrix of the system at a set of times, $\hrho(t_i)$. Gradient-based approaches to optimize the control parameters rely on computing the derivative of the cost function with respect to each parameter, $\dd C / \dd \theta$. To do so, we apply the adjoint state method~\cite{pontryagin1962mathematical} to open quantum systems. In this context, the adjoint state is defined as $\hphi(t) = \dd C / \dd \hrho(t)$, and represents how a change in the density matrix at time $t$ modifies the cost function. For open quantum systems under the usual Born-Markov approximations~\cite{gardiner2004quantum}, the evolution of the density matrix is governed by a Lindblad master equation ($\hbar = 1$),
\begin{equation}\label{eq:mastereq}
    \frac{\dd\hrho}{\dd t} = \mathcal{L}\hrho \equiv -i [\H, \hrho] + \sum_k \mathcal{D}[\L_k]\hrho,
\end{equation}
where $\H$ is the system Hamiltonian, $\L_k$ are jump operators, and $\mathcal{D}[\L]\hrho = \L \hrho \L^\dagger - \{\L^\dagger \L, \hrho \} / 2$. The adjoint state is then subject to a dual ordinary differential equation~\cite{supmat},
\begin{equation}\label{eq:adjointeq}
    \frac{\dd\hphi}{\dd t} = - \mathcal{L}^\dagger \hphi \equiv -i [\H, \hphi] - \sum_k \mathcal{D}^\dagger[\L_k]\hphi,
\end{equation}
where $\mathcal{D}^\dagger[\L]\hphi = \L^\dagger \hphi \L - \{\L^\dagger \L, \hphi \} / 2$. This equation can be integrated numerically over the time interval of interest $[t_0, t_n]$ with initial condition $\hphi(t_n) = \partial C / \partial \hrho(t_n)$, computed analytically if a closed form is available, or directly through automatic differentiation. Notably, the overall minus sign in \cref{eq:adjointeq} ensures numerical stability of the integration by generating contracting dynamics in reverse time. The derivative of the cost function with respect to the problem parameters is given by
\begin{equation}\label{eq:gradeq}
    \frac{\dd C}{\dd\theta} = \frac{\partial C}{\partial \theta} - \int_{t_n}^{t_0} \partial_\theta \Tr[\hphi^\dagger(t) \mathcal{L}(t, \theta) \hrho(t)] \dd t\,.
\end{equation}
This integral is straightforward to compute using the density matrix and adjoint state at each time $t \in [t_0, t_n]$, as obtained from~\cref{eq:mastereq,eq:adjointeq}. In particular, the partial derivative with respect to $\theta$ can be easily computed from automatic differentiation of the adjoint state equation by noting that
\begin{equation}
\partial_\theta \Tr[\hphi^\dagger \mathcal{L} \hrho] = - \Tr[ \partial_\theta (\dd\hphi / \dd t)^\dagger \hrho],
\end{equation}
which has the form of a vector-Jacobian product.

\begin{figure}[!t]
         \centering
         \includegraphics[width = \columnwidth]{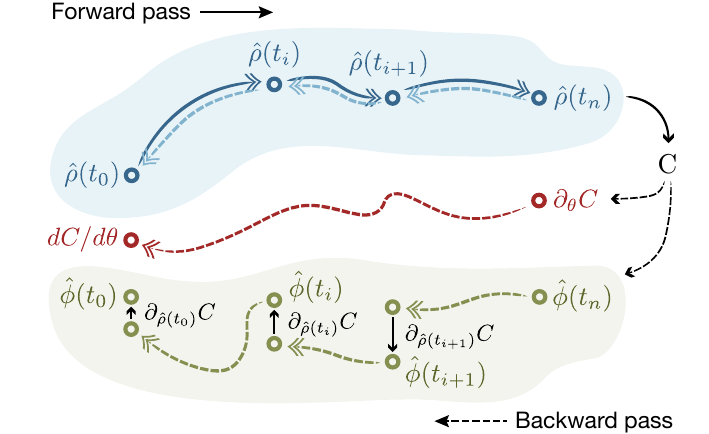}
\label{fig:adjointschematic}
\vspace{-0.5cm}
\caption{
    Adjoint state quantum optimal control.\ In the forward pass, the master equation is integrated and checkpointed for several time points (dark blue). In the backward pass, the density matrix $\hat\rho$ is recomputed in reverse time (light blue) together with the adjoint state $\hat\phi$ (green) and with the gradients (red) of the cost function $C$. When a checkpoint is reached, the density matrix is restored to its forward time trajectory, and the adjoint state updated with the corresponding cost function gradient.
}
 \end{figure}

The QOC optimization is illustrated in~\cref{fig:adjointschematic} and proceeds in two steps. First, the forward pass consists in using the initial set of parameters (e.g.~a sequence of discrete pulses) to numerically integrate the master equation from $t_0$ to $t_n$ while saving the density matrix at each time $t_i$ of interest. The cost function $C(\theta, \hrho(t_0), \ldots, \hrho(t_n))$ is then evaluated. To lower the memory footprint, the cost function can also be evaluated on the fly during the forward pass such that only a single density matrix needs to be stored. In a second step, the backward pass, both the master and adjoint equations are simultaneously integrated in reverse time, starting from $t=t_n$. During this process, the integral of~\cref{eq:gradeq} is iteratively evaluated, such as to obtain the entire gradients $\dd C / \dd \theta$ once the backpropagation is finished. Having access to the gradients of the cost function, we can now iteratively update the control parameters using standard optimization algorithms~\cite{broyden1970convergence,kingma2014adam,robbins1951stochastic}.

We emphasize how each density matrix (blue) is computed twice: once during the forward pass, and once during the backward pass. This enables a low memory footprint for the overall scheme, with at most a single density matrix and adjoint state needed to be stored at any given moment. The memory footprint of the method thus scales as $\mathcal{O}(N^2)$ with $N$ the Hilbert space dimension. This is in stark contrast with methods based on automatic differentiation~\cite{leung2017speedup}, for which the density matrix needs to be stored at each time point of the numerical integration, thus scaling as $\mathcal{O}(n N^2)$, with $n$ the number of numerical integration steps~\cite{lu2024optimal}. Such memory requirements can quickly become prohibitive, even for open quantum systems of intermediate sizes, $N\gtrsim 100$.

Note that this large gain in memory comes at the cost of trading off some numerical runtime. Overall, the scheme requires the integration of four differential equations in total~\cite{supmat}, against only two for automatic differentiation. In addition, the reverse time integration of~\cref{eq:mastereq} can be numerically unstable due to the expansive dynamics of the system. This can however be fully resolved with checkpointing the quantum states during the forward pass~\cite{narayanan2022reducing}, thus effectively trading back some memory for numerical stability. In practice, checkpointing at the time scale of the largest dissipation operator is sufficient to ensure numerical stability without adding significant complexity.

We have implemented this optimization scheme using PyTorch~\cite{paszke2019pytorch}, taking advantage of its automatic differentiation capabilities and GPU support. This framework allows us to run optimization problems for open quantum system with hundreds of parameters, arbitrary cost functions, and for Hilbert space dimensions of up to $N \sim \num{5000}$ while running on a single GPU with 24~GB of memory. Our code is available through the \texttt{dynamiqs} open-source library~\cite{dynamiqs}, simplifying replication of this work and its application to various QOC problems. We now demonstrate the usefulness of this method by optimizing readout and reset of a transmon, two operations that inherently rely on dissipation.

\begin{figure*}
    \centering
    \includegraphics[width = \textwidth]{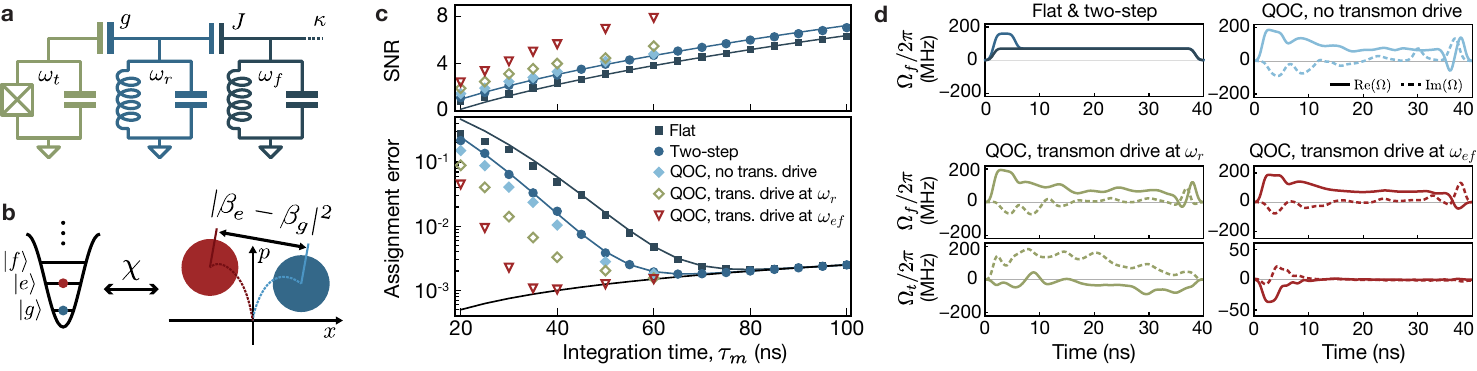}
\vspace{-0.5cm}
\caption{\label{fig:readout}
    (a) Lumped-element model of a transmon coupled to a readout resonator and Purcell filter. The transmon and filter are driven, and the filter output field is measured through its transmission line.\ (b) Dispersive readout of a transmon. The mean field in the filter depends on the transmon state. SNR of readout increases with the integrated difference of mean fields.\ (c) SNR and assignment error of transmon readout for several drive envelopes: a flat envelope, a \SI{4}{\nano\second} two-step envelope~\cite{walter2017rapid}, and optimized envelopes (QOC) with optional additional drives on the transmon at frequencies $\omega_{d,t} \simeq \omega_r$ (green) and $\omega_{d,t} \simeq \omega_{ef}$ (red). The flat and two-step data points are fitted according to \cref{eq:snrfit}. The black line shows the $T_1$ limit given by $\tau_m / 2 T_1$.\ (d) Reference and optimized pulse envelopes at \SI{40}{\nano\second} of integration time.
}
\end{figure*}

\lettersection{Transmon model}
Let us consider the experimentally realistic model depicted in~\cref{fig:readout}(a) of a transmon coupled to a readout resonator and Purcell filter~\cite{blais2021circuit}
\begin{equation}
    \frac{\dd\hrho}{\dd t} = -i[\H, \hrho] + \gamma \mathcal{D}[\b]\hrho + \kappa \mathcal{D}[\f]\hrho,
\end{equation}
with transmon relaxation rate $\gamma$ and filter relaxation rate $\kappa$, and where
\begin{equation}\label{eq:readout-hamil}
    \begin{split}
        \H &= 4 E_C \n_t - E_J \cos(\Hat{\varphi}_t) + \omega_r \adag \a + \omega_f \fdag \f \\
        &- i g \n_t (\a - \adag) - J (\a - \adag)(\f - \fdag) \\
        &+ \Omega_t \n_t \sin(\omega_{d,t} t) - i\Omega_f (\f - \fdag) \sin(\omega_{d,f} t).
    \end{split}
\end{equation}
The first two terms denote the free transmon Hamiltonian with charging energy $E_C$ and Josephson energy $E_J$, with $\n_t$ and $\Hat{\varphi}_t$ the charge and phase operators, and with $\b$ the corresponding annihilation operator in the diagonal basis. The resonator and filter modes are denoted by $\a$ and $\f$, with respective frequencies $\omega_r$ and $\omega_f$. These three modes are capacitively coupled in series with coupling strengths $g \gg J$. The system can be driven using a capacitive coupling either through the transmon with a microwave pulse at frequency $\omega_{d,t}$ and envelope $\Omega_t(t)$, or through the Purcell filter at frequency $\omega_{d,f}$ and envelope $\Omega_f(t)$.

For numerical simulation of this model, we first diagonalize the free transmon Hamiltonian and identify the lowest energy eigenstates. We also diagonalize the resonator-filter subsystem yielding two normal modes, each coupled to the transmon. Finally, we apply the rotating-wave approximation (RWA) on couplings and drives. This allows for larger numerical time steps by eliminating fast oscillating dynamics thereby simplifying master equation integration. However, this also implies that not all of the chaotic or transmon ionization dynamics are captured~\cite{cohen2022reminiscence,shillito2022dynamics,dumas2024unified}. To avoid probing these regimes, we limit the maximum amplitudes of control drives, e.g.~to $\SI{200}{\mega\hertz}$ for transmon readout.

We use typical device parameters corresponding to a critical photon number of $\bar{n}_\mathrm{crit} = {(\Delta / 2g)}^2 = 16$~\cite{blais2004cavity} and dispersive rates of $\chi / 2\pi = \SI{3.8}{\mega\hertz}$ and $\SI{8.1}{\mega\hertz}$ with the lower and higher normal modes, respectively. The filter loss rate is $\kappa / 2\pi = \SI{30}{\mega\hertz}$ and the transmon relaxation time is $T_1 = \SI{20}{\micro\second}$. The remaining system parameters can be found in~\cite{supmat}.

\lettersection{Transmon readout}
Readout of transmon qubits is realized through the dispersive coupling to a resonator~\cite{blais2004cavity}. In this case, the resonator frequency is shifted by the average occupancy in the transmon, and can be measured by driving the resonator at its bare frequency and monitoring the output field, see \cref{fig:readout}(b). In the presence of a Purcell filter, either normal mode of the hybridized resonator-filter subsystem can be used for readout~\cite{swiadek2023enhancing}.

The metric we use to maximize the measurement fidelity is the SNR. Accounting for optimal weighting functions~\cite{ryan2015tomography}, it reads~\cite{bultink2018general}
\begin{equation}\label{eq:snr-bultink}
    \mathrm{SNR}(\tau_m) = \sqrt{2 \eta \kappa\int_0^{\tau_m} \dd t\, |\beta_e(t) - \beta_g(t)|^2 },
\end{equation}
where $\eta \in [0, 1]$ is the measurement efficiency, $\tau_m$ is the readout integration time, and $\beta_{e/g} = \Tr\big[\f \hrho_{g/e}\big]$ is the average field value in the filter mode, with $\hrho_{g/e}$ the density matrix obtained after initializing the transmon in the $\ket{g/e}$ state. To obtain results that can be compared to experiments, we use $\eta=0.6$~\cite{walter2017rapid}. The optimization objective is to maximize the SNR, and thus maximize the distance between the pointer states $|\beta_e - \beta_g|$ in the shortest possible time. Further assuming that the pointer states $\beta_{g,e}$ are Gaussian, one can link the SNR and the transmon lifetime to the readout assignment error~\cite{gambetta2007protocols,supmat}.

To optimize the transmon readout, we discretize the control pulse envelopes $\Omega(t)$ with \SI{1}{\nano\second} time bins and use a $\SI{250}{\mega\hertz}$ gaussian filter to interpolate between these pixels during numerical integration and to model realistic experimental distortions~\cite{motzoi2011optimal}. In addition to the discretized drive amplitudes, the optimization parameters $\theta$ include the carrier frequency $\omega_d$ of each drive. Contrary to the drive amplitudes, the latter are kept constant throughout the pulse duration, in accordance with typical experiments. The cost function used to optimize the transmon readout is principally composed of the SNR of~\cref{eq:snr-bultink}, with additional cost terms constraining the control pulses in order to regularize the optimization and avoid out-of-model dynamics. For example, we limit the number of photons in the hybridized resonator-filter modes, penalize unwanted transitions to higher excited transmon states, and limit the maximal available pulse amplitudes. The full cost function is detailed in~\cite{supmat}.
We perform gradient descent using Adam~\cite{kingma2014adam} and use the adjoint state method previously described to compute gradients.

\Cref{fig:readout}(c) shows the SNR and the assignment error as a function of the integration time $\tau_m$ obtained by our approach, and panel (d) shows the corresponding pulse envelopes for $\tau_m = \SI{40}{\nano\second}$ optimizations. As a point of comparison, we first consider the two non-optimized reference pulses labelled `flat' and `two-step'. The former consists of a constant pulse with \SI{2}{\nano\second} ramp-up and ramp-down times (dark blue squares), and the latter of a two-step pulse meant to rapidly populate the readout mode (blue circles)~\cite{walter2017rapid}. In both cases, the amplitude is calibrated to reach $\bar{n} = \bar{n}_\mathrm{crit}$ photons in the steady state. The SNR versus $\tau_m$ for these two pulses is fitted with the function (full blue and dark blue lines)~\cite{blais2021circuit}
\begin{equation}\label{eq:snrfit}
    \mathrm{SNR}(\tau_m) = \alpha \sqrt{2 \eta \kappa} \left(\sqrt{\tau_m} - \sqrt{\tau_{m,0}}\right),
\end{equation}
where $\alpha = 2 |\Omega_f \sin(2 \phi)| / \kappa$ is the effective resonator displacement in the steady state, with $\phi = \arctan(2\chi / \kappa)$ and $\chi$ the dispersive shift obtained from exact diagonalization of \cref{eq:readout-hamil}. In this expression, $\sqrt{\tau_{m,0}}$ accounts for an initial delay for the resonator to populate, and is numerically fitted to $\tau_{m, 0} = \SI{19}{\nano\second}$ and $\tau_{m,0} = \SI{13}{\nano\second}$ for the flat and two-step pulses, respectively. As the integration time increases, the SNR (assignment error) of both references pulses increase (decrease), up until the transmon $T_1$ limit is reached (solid black line). Minimum assignment errors of $2.1\times10^{-3}$ and $1.8\times10^{-3}$ are obtained at $\SI{80}{\nano\second}$ and $\SI{65}{\nano\second}$ respectively. This is similar performance to state-of-the-art readout experiments~\cite{walter2017rapid,sunada2022fast,swiadek2023enhancing,sunada2024photon}, as expected from our choice of realistic experimental parameters. Our objective is now to obtain smaller assignment errors in shorter measurement times.

The light blue symbols in \cref{fig:readout}(c) are obtained by optimizing the pulse envelope and drive frequency using our QOC approach. The gain is modest and mainly limited by the dispersive coupling with the transmon. Interestingly, the optimized pulses follow a two-step-like shape with a strong initial drive and a weaker subsequent drive, see panel (d). We attribute the small oscillations in the envelope to the rotational gauge freedom of the resonators, which the optimizer is arbitrarily choosing.

Significant improvements are, however, obtained by adding a drive on the transmon concurrently to the readout drive on the resonator. Interestingly, the optimizer converges on two distinct frequencies for the transmon drive. The first strategy found by the optimizer is to drive the transmon at a frequency close to the resonator frequency (green symbols). In that case, the assignment errors decreases faster with integration time than with the above approaches, leading to a minimal assignment error of $1.6\times10^{-3}$ at $\SI{60}{\nano\second}$. The effectiveness of this optimized readout strategy stems from the fact that driving the qubit at the resonator frequency creates a longitudinal-like interaction that can be combined with the usual dispersive interaction to improve readout, as demonstrated in Refs.~\cite{ikonen2019qubit,touzard2019gated,munoz2023qubit}.

The second strategy found by the optimizer employs a transmon drive at the (ac-Stark shifted) $\ket{e}$-$\ket{f}$ transition frequency (red symbols). Given that the cavity response differs more significantly between the transmon states $\ket g$ and $\ket f$ than between $\ket g$ and $\ket e$~\cite{blais2021circuit}, transferring population into the $\ket f$ state leads to a significant improvement of the assignment error, which reaches $1.0\times10^{-3}$ in $\SI{40}{\nano\second}$. Interestingly, this shelving approach has already been used to improve readout in circuit QED~\cite{mallet2009,danjou2017enhancing,elder2020,chen2023transmon}. There, a $\pi$-pulse between $\ket e$ and $\ket f$ is applied to the transmon followed by the measurement drive.  In contrast, the optimized strategy found here applies the $\pi$-pulse while the cavity is loaded with measurement photons leading to a considerable reduction in the measurement time, see~\cref{fig:readout}(d). This is possible because the optimizer accounts for the time-dependent ac-Stark shift. The optimized $\pi$-pulse features a DRAG-like envelope~\cite{motzoi2009simple} and achieves a gate fidelity over 99~\% in less than \SI{10}{\nano\second}, even while the readout mode is being strongly driven. Importantly, we note that this approach could achieve significantly higher fidelities by increasing the modest transmon lifetime of $\SI{20}{\micro\second}$ used here, as shown by the high SNR in~\cref{fig:readout}(c).

\begin{figure}
    \centering
    \includegraphics[width = \columnwidth]{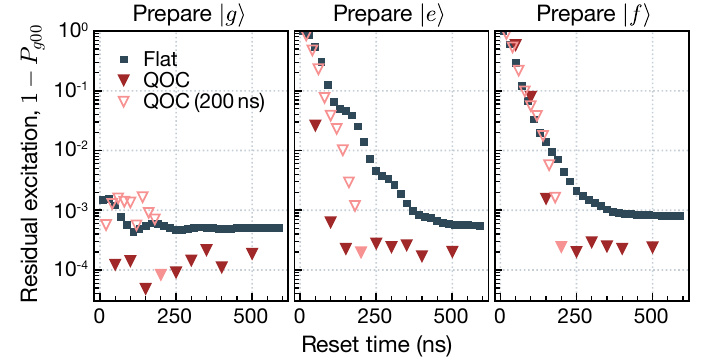}
\vspace{-0.5cm}
\caption{\label{fig:reset}
    Residual excitation out of $\ket{g00}$ after f0-g1 reset for a calibrated pulse (blue) and an optimized pulse (red) for a system prepared in $\ket{g00}$, $\ket{e00}$, and $\ket{f00}$. Each filled red marker corresponds to a different optimization. Lighter hollow markers illustrate the population dynamics at shorter times for the optimized pulse with \SI{200}{\nano\second} duration.
}
\end{figure}

\lettersection{Transmon reset}
As a second demonstration of the adjoint state method, we consider the optimization of the f0-g1 reset of a transmon~\cite{pechal2014microwave,zeytinouglu2015microwave,magnard2018fast}. This is an all-microwave reset protocol based on a Raman transition between states $\ket{f00}$ and $\ket{g01}$. For the ket $\ket{ijk}$, $\ket{i}$ stands for the qubit state and $\ket{jk}$ the resonator-filter normal modes. Given the large photon loss rate of the filter, the state $\ket{g01}$ quickly decays to $\ket{g00}$, thus ensuring a fast reset of the transmon $\ket{f}$ state. An additional drive at the $\ket{e}$-$\ket{f}$ transition frequency allows to reset both $\ket{e}$ and $\ket{f}$ states of the transmon. We use the adjoint-state method to find optimal controls for both the f0-g1 and e-f drives simultaneously, in a similar fashion as for optimizing the readout. The cost function is now principally maximizing the transmon population in the $\ket{g00}$ state at the end of the protocol, along with smaller contributions for regularizing the pulses, see~\cite{supmat} for details.

The results of the reset optimization are summarized in \cref{fig:reset}. The three panels show the residual excitation out of $\ket{g00}$ against the reset time for a reference flat pulse (blue) and an optimized pulse (red) for different initial transmon states. The reference pulse is composed of two constant drives at the f0-g1 and e-f transitions, where amplitudes and frequencies are calibrated numerically in a similar fashion to what is done in experiments, see~\cite{supmat}. The QOC pulse is obtained by optimizing the carrier frequency and envelopes of both drives, for several total reset times. The optimized pulses show significant improvement over the reference, with a residual excitation of less than 0.05\:\% at \SI{100}{\nano\second} (\SI{200}{\nano\second}) for the $\ket{e}$ ($\ket{f}$) state preparation. Note that this delay in the $\ket{f}$ reset time is due to a larger relative weight for the reset of $\ket{e}$ chosen in the cost function, and could be adjusted to achieve the most experimentally relevant reset scheme. This represents a notable improvement over the reference pulses, which reach a steady state after more than \SI{300}{\nano\second} with larger residual excitations of about 0.07\:\%. Our results also favourably compare to state-of-the-art experimental realizations of this protocol that reach 1.7\:\% residual excitations in \SI{100}{\nano\second}~\cite{sunada2022fast}, or 0.3\:\% in \SI{300}{\nano\second}~\cite{magnard2018fast}.

\lettersection{Conclusion}
We obtained a fully general framework to optimize open quantum system dynamics in large Hilbert spaces by combining the adjoint state method and reverse-time backpropagation.
We have demonstrated the applicability of this method to complex open-system optimization problems using the example of superconducting transmon readout and reset. 
We stress that our method can readily be applied to optimizing a wide range of quantum control problems where the dissipative dynamics play a significant role such as reservoir (dissipation) engineering~\cite{poyatos1996quantum,harrington2022engineered}, autonomous QEC~\cite{royer2020stabilization,gertler2021protecting}, leakage-reduction units~\cite{aliferis2007lru}, quantum cooling, and more.
We encourage readers to apply this framework on their own optimal control problems using the open-source library \texttt{dynamiqs}~\cite{dynamiqs}.

\lettersection{Acknowledgments}
We sincerely thank Alain Sarlette, Ross Shillito, Cristobal Lled\'o, Pierre Guilmin and Adrien Bocquet for useful discussions. R.G. extends his gratitude to Alain Sarlette for helping organise and support this long-term academic exchange. This work was supported by grant ANR-18-CE47-0005, NSERC, the Canada First Research Excellence Fund, and the Minist\`ere de l’\'Economie et de l’Innovation du Qu\'ebec. The numerical simulations were performed using HPC resources at Institut quantique and Inria Paris.

\let\oldaddcontentsline\addcontentsline
\renewcommand{\addcontentsline}[3]{}
\bibliography{bibliography}
\let\addcontentsline\oldaddcontentsline


\clearpage

\onecolumngrid
{\center \bf \large Supplemental Material for "Optimal control in large open quantum systems --\\the case of transmon readout and reset"\\}
{\center Ronan Gautier,\textsuperscript{1,\,2,\,3} \'Elie Genois,\textsuperscript{1} and Alexandre Blais\textsuperscript{1,\,4}\\\vspace*{-0.2cm}}
{\center \small \textsuperscript{1}\textit{Institut Quantique and D\'epartement de Physique,\\Universit\'e de Sherbrooke, Sherbrooke, QC, Canada}\\\vspace*{-0.3cm}}
{\center \small \textsuperscript{2}\textit{Laboratoire de Physique de l'\'Ecole Normale Sup\'erieure, Inria, ENS,\\ Mines ParisTech, Universit\'e PSL, Sorbonne Universit\'e, Paris, France}\\\vspace*{-0.3cm}}
{\center \small \textsuperscript{3}\textit{Alice \& Bob, Paris, France}\\\vspace*{-0.3cm}}
{\center \small \textsuperscript{4}\textit{Canadian Institute for Advanced Research, Toronto, ON, Canada}\\\vspace*{1.cm}}
\twocolumngrid

\setcounter{equation}{0}
\setcounter{figure}{0}
\setcounter{table}{0}
\setcounter{page}{1}
\setcounter{section}{0}

\makeatletter
\renewcommand{\theequation}{S\arabic{equation}}
\renewcommand{\thefigure}{S\arabic{figure}}
\renewcommand{\thesection}{S\arabic{section}}
\renewcommand{\thetable}{S\arabic{table}}
\renewcommand{\thealgocf}{S\@arabic\c@algocf}

\tableofcontents

\section{Adjoint state method for open quantum systems}\label{app:adjointmethod}
\subsection{Adjoint state master equation}\label{app:proof}

In this section, we prove that the adjoint state follows the differential equation Eq.~(2) shown in the main text, following a similar derivation as in Ref.~\cite{chen2018neural}. To formalize this derivation, we introduce the double-ket and double-bra notation by making the changes $\hrho \leftrightarrow \kket{\rho}$ and $\hphi^\dagger \leftrightarrow \bbra{\phi}$~\cite{albert2018lindbladians}. Starting from a general Lindblad master equation
\begin{equation}\label{eq:sm-mastereq}
    \frac{\dd\kket{\rho}}{\dd t} = \mathcal{L}\kket{\rho} \equiv \left(-i [\H, \cdot\,] + \sum_k \mathcal{D}[\L_k] \right)\kket{\rho},
\end{equation} 
we can write that for an $\varepsilon$ change in time, the evolved quantum state takes the form
\begin{equation}
    \begin{split}
        \kket{\rho(t + \varepsilon)} &= \kket{\rho(t)} + \int_t^{t+\varepsilon} \mathcal{L}(\tau, \theta) \kket{\rho(\tau)} \dd \tau \\
        &= \kket{\rho(t)} + \varepsilon \mathcal{L}(t, \theta) \kket{\rho(t)} + \mathcal{O}(\varepsilon^2).
    \end{split}
\end{equation}
Using the definition of the adjoint state and applying the chain rule, we find
\begin{equation}\label{eq:chainrule}
\begin{split}
    \bbra{\phi(t)} &= \frac{\dd C}{\dd \kket{\rho(t)}} = \frac{\dd C}{\dd \kket{\rho(t+\varepsilon)}} \frac{\dd \kket{\rho(t + \varepsilon)}}{\dd \kket{\rho(t)}} \\
    &= \bbra{\phi(t + \varepsilon)} \left( 1 + \varepsilon \mathcal{L}(t, \theta) + \mathcal{O}(\varepsilon^2) \right).
\end{split}
\end{equation}
Then, the proof of the adjoint state master equation follows directly from the definition of the derivative,
\begin{equation}
    \begin{split}
        \frac{\dd \bbra{\phi(t)}}{\dd t} &= \lim_{\varepsilon \rightarrow 0} \frac{\bbra{\phi(t+\varepsilon)} - \bbra{\phi(t)}}{\varepsilon} \\
        &= - \lim_{\varepsilon \rightarrow 0} \left(\bbra{\phi(t+\varepsilon)} \mathcal{L}(t, \theta) + \mathcal{O}(\varepsilon^2) \right) \\
        &= - \bbra{\phi(t)} \mathcal{L}(t, \theta)
    \end{split}
\end{equation}
or equivalently, taking the hermitian conjugate of this equation, we get the following result thus completing the proof
\begin{equation}\label{eq:sm-adjointeq}
    \frac{\dd\kket{\phi(t)}}{\dd t} = - \mathcal{L}^\dagger(t, \theta) \kket{\phi(t)}.
\end{equation}

\subsection{Explicit expression of gradients}

To derive an explicit expression for $\dd C / \dd \theta$, we also follow Ref.~\cite{chen2018neural} and introduce an augmented density matrix $\kket{\rho_{aug}(t)} = [\kket{\rho(t)}, \theta(t)]^T$ which treats the parameters as an additional state to integrate. In practice, parameters are constant throughout the integration, but this fictitious time-dependence will allow us to isolate the sensitivity of the cost function to an infinitesimal change of parameters at each point in time. The augmented density matrix follows the differential equation
\begin{equation}
    \frac{\dd \kket{\rho_{aug}(t)}}{\dd t} = \mathcal{L}_{aug}(t, \theta) \kket{\rho_{aug}(t)} = 
    \begin{pmatrix}
        \mathcal{L}(t, \theta) \kket{\rho(t)} \\
        0
    \end{pmatrix}.
\end{equation}
We can also introduce the augmented adjoint state, $\bbra{\phi_{aug}(t)} = [\bbra{\phi(t)}, \dd C / \dd \theta(t)]^T$, where the parameter sensitivity $\dd C / \dd\theta(t)$ represents the cost function gradient with respect to the parameters at a given time. We then follow a similar derivation as in the previous section. The main difference comes in the application of the chain rule of~\cref{eq:chainrule} which now yields,
\begin{equation}
    \begin{split}
        \bbra{\phi_{aug}(t)} &= \frac{\dd C}{\dd \kket{\rho_{aug}(t)}} = \frac{\dd C}{\dd \kket{\rho_{aug}(t + \varepsilon)}} \frac{\dd \kket{\rho_{aug}(t + \varepsilon)}}{d \kket{\rho_{aug}(t)}} \\
        &= \bbra{\phi_{aug}(t+ \varepsilon)}\left(1 + \varepsilon
        \mathcal{M}(t, \theta) + \mathcal{O}(\varepsilon^2)\right)
    \end{split}
\end{equation}
where
\begin{equation}
    \begin{split}
        \mathcal{M}(t, \theta) &= \frac{\dd \left(\mathcal{L}_{aug}(t, \theta) \kket{\rho_{aug}(t)}\right)}{\dd \kket{\rho_{aug}(t)}} \\
        &=
        \begin{pmatrix}
            \mathcal{L}(t, \theta) & \partial_{\theta} \mathcal{L}(t, \theta)\kket{\rho(t)} \\
            0 & 0
        \end{pmatrix},
    \end{split}
\end{equation}
and where $\partial_{\theta}$ denotes the partial derivative with respect to $\theta$. Similarly as before, we can rearrange this expression, reinsert it in the definition of the derivative and take the limit of $\varepsilon \rightarrow 0$. Then, selecting only the second block of the resulting differential equation yields
\begin{equation}
    \frac{\dd}{\dd t} \left(\frac{\dd C}{\dd \theta(t)}\right) = - \partial_{\theta} \bbra{\phi(t)} \mathcal{L}(t, \theta) \kket{\rho(t)}
\end{equation}
We have thus obtained a straightforward differential equation on the parameter sensitivity. Integrating this equation from $t=0$ to $T$, with the initial condition that $\dd C / \dd \theta(T) = \partial C / \partial \theta$ yields
\begin{equation}\label{eq:pgrad}
    \begin{split}
        \frac{\dd C}{\dd \theta} &= \frac{\partial C}{\partial \theta} - \int_T^0 \partial_{\theta} \bbra{\phi(t)} \mathcal{L}(t, \theta) \kket{\rho(t)} \dd t \\
        &=\frac{\partial C}{\partial \theta} - \int_T^0 \partial_{\theta} \bbrakket{\mathcal{L}^\dagger(t, \theta) \phi(t)}{\rho(t)} \dd t,
    \end{split}
\end{equation}
which, using that $\bbra{\phi}\mathcal{L}\kket{\rho} = \Tr[\hphi^\dagger \Liouv \hrho]$, is Eq.~(3) in the main text. We can also get the derivative of the loss function with respect to the integration time $T$ from a simple application of the chain rule
\begin{equation}\label{eq:tgrad}
    \frac{\dd C}{\dd T} = \frac{\dd C}{\dd \kket{\rho(T)}} \frac{\dd \kket{\rho(T)}}{\dd T} = \bbra{\phi(T)}\mathcal{L}(T, \theta) \kket{\rho(T)}.
\end{equation}
Finally, between \cref{eq:sm-adjointeq,eq:pgrad,eq:tgrad}, we obtain the gradients of the loss function with respect to all relevant objects involved in the computation, thus allowing one to perform gradient descent for the desired parameters from arbitrary cost functions. In addition, note that \cref{eq:pgrad} is exact, such that the accuracy of the computed gradients is only limited by numerical errors arising from the master equation or adjoint master equation integration. 

\subsection{Interpretation of the adjoint state in the quantum setting}

In the general setting of ordinary differential equations, the adjoint state can be though of as the continuous-time analog of the chain rule. However, in the quantum setting, a deeper interpretation of this object can arise.

A helpful example to gain intuition into this is to consider a cost function in the form of an expectation value, i.e. $C(\hrho(t_n)) = \Tr[\hO \hrho(t_n)]$ with $\hO$ an arbitrary operator. In this case, we have that $\hphi(t_n) = \hO$ and the adjoint state differential equation in reverse time reads
\begin{equation}
    \frac{\dd\hat\phi}{\dd(-t)} = \mathcal{L}^\dagger \hat\phi = i [\hat H, \hat\phi] + \sum_k \mathcal{D}^\dagger[\hat L_k] \hat \phi
\end{equation}
which is exactly the standard Lindblad master equation in the Heisenberg picture~\cite{breuer2002theory}. As a consequence, we see that the adjoint state is exactly the Heisenberg representation of the operator of interest, $\hphi = \hO_H$.

In general, the cost function is not necessarily in the form of an expectation value, or even linear in the density matrices involved, but one could still interpret the adjoint state as a generalized operator in the Heisenberg picture.

\SetCommentSty{textsl}
\begin{algorithm}[t!]
    \caption{Adjoint state method for open quantum systems.}
    \label{alg:adjointmethod}
    $t \gets t_0$\;
    $\rho \gets \rho_0$\;
    $L_\rho \gets \{\rho_0\}$\tcp*[r]{checkpoint storage list (optional)}
    \For(\tcp*[f]{forward pass}){$t_i \in \{t_1, \ldots, t_n\}$}{
        \While{$t < t_i$}{
            $\rho, \delta t \gets \textsf{ODEstep}(\mathcal{L}, t, \rho, \theta)$\;
            $t \gets t + \delta t$
        }
        $L_\rho \gets L_\rho + \{\rho\}$\;
    }
    $C \gets \textsf{CostFn}(\theta, L_\rho)$\;
    $\phi \gets \textsf{AutoDiff}(\textsf{CostFn}, C, L_\rho[n])$\;
    $\nabla \gets \textsf{AutoDiff}(\textsf{CostFn}, C, \theta)$\;
    \For(\tcp*[f]{backward pass}){$t_i \in \{t_{n-1}, \ldots, t_0\}$}{
        \While{$t > t_i$}{
            $\{\rho, \phi\}, \delta t \gets \textsf{ODEstep}\!\left(\{-\mathcal{L}, \mathcal{L}^\dagger\}, t, \{\rho, \phi\}, \theta \right)$\;
            $\delta \phi \gets \textsf{AutoDiff}(\textsf{ODEstep}, \phi, \theta)$\;
            $\nabla \gets \nabla - \mathrm{Tr}\!\left[\delta\phi \cdot \rho \right]$\;
            $t \gets t - \delta t$\;
        }
        $\phi \gets \phi + \textsf{AutoDiff}(\textsf{CostFn}, C, L_\rho[i])$\;
        $\rho \gets L_\rho[i]$\;
    }
    \Return{$\nabla$}
\end{algorithm}

\subsection{Practical implementation}

The full implementation in Python of the approach described in the main text is available on the freely available open-source software dynamiqs~\cite{dynamiqs}.
To summarize the structure of a numerical implementation of the adjoint state method for open quantum systems, we present a pseudo-code implementation in~\cref{alg:adjointmethod}. It involves a function $\textsf{ODEstep}(f, t, y, \theta)$ that computes a single step of the integration of a linear ordinary differential equation, $\dot{y} = f(t, \theta) y$, at time $t$. The $\textsf{CostFn}(\theta, L_\rho)$ function computes the cost function from the parameters $\theta$ and a list of density matrices $L_\rho$. It can also be computed on the fly in the \emph{for} loop to avoid storing these matrices. Finally, the $\textsf{AutoDiff}(g, x_{out}, x_{in})$ function computes the partial derivative $\partial x_{out} / \partial x_{in}$ through automatic differentiation of $g$, where $x_{in}$ and $x_{out}$ are given inputs and outputs of the function $g$. Note that lines 17 and 18 can be merged in a single computation of a vector-Jacobian product to avoid storage of $\delta\phi$.

\begin{table}[!t]
    \centering
    \def\arraystretch{1.4}
    \begin{tabular*}{\columnwidth}{@{\extracolsep{\fill}} lcc}
    \hline \hline
    & Runtime & Memory usage \\ \hline \\[-1.5em]
    Automatic differentiation & $\mathcal{O}(2\mu n N^3)$ & $\mathcal{O}(\mu n N^2)$ \\
    Adjoint state method & $\mathcal{O}(4\mu n N^3)$ & $\mathcal{O}(\mu N^2)$ \\ \hline\hline
    \end{tabular*}
    \caption{Runtime and memory usage scaling of integration and differentiation of the Linblad master equation for automatic differentiation and for the adjoint state method with reverse-time backpropagation. The scaling is given as a function of the Hilbert space size $N$, the number of time steps in the numerical integration $n$, and the number of matrix-matrix operations for each step $\mu$. }
    \label{tab:memorycost}
\end{table}

\subsection{Time and memory cost}

In \cref{tab:memorycost}, we provide a comparison of the runtime and memory usage scaling for integration (forward pass) and differentiation (backward pass) of the Lindblad master equation for automatic differentiation and for the adjoint state method with reverse-time backpropagation. The scaling is given as a function of the Hilbert space size $N$, the number of integration time steps $n$, and the number of matrix-matrix operations at each step $\mu$. The number of steps $n$ typically scales linearly in the simulation time, and inversely with the timescale of the fastest dynamics in the system under study. The number of matrix-matrix operations at each step $\mu$ depends on the integration method used (e.g. Dormand-Prince of order 4/5~\cite{dormand1980family}) and on the number of jump operators involved in the master equation.

For automatic differentiation, the runtime scales according to $\mathcal{O}(2\mu n N^3)$ where $N^3$ is the cost of a single matrix-matrix operation, $\mu n$ the total number of such operations per pass, and $2$ being the forward and backward pass respectively. Importantly, the memory usage scales as $\mathcal{O}(\mu n N^2)$ where $N^2$ is the number of elements in a single density matrix, and $\mu n$ is the number of such matrices that are stored during the forward pass to be reused in the backward pass. This is therefore a linear scaling in the number of time steps, which can quickly be prohibitive even for moderate system sizes and number of time steps.

For the adjoint state method with reverse-time backpropagation, as presented in the main text, the runtime scales according to $\mathcal{O}(4\mu n N^3)$. This additional factor of $2$ compared to automatic differentiation arises from the fact that 3 differential equations need to be solved during the backward pass: the reverse-time Lindblad equation, the reverse-time adjoint state master equation, and the automatic differentiation of the latter equation at each step for gradient computation (see Eqs.~(3-4) in the main text). However, the memory scaling is drastically reduced with a scaling in $\mathcal{O}(\mu N^2)$, corresponding to the memory cost of storing the computational graph of the adjoint state during each step of the backward pass. This graph must only be stored for a single step at a time, thus the constant scaling with $n$.

\begin{table*}[!t]
  \centering
  \def\arraystretch{1.4}
  \begin{tabular*}{\linewidth}{@{\extracolsep{\fill}} lccc}
    \hline \hline
    \textbf{Description} & \textbf{Analytical formula} & \textbf{Weight} & \textbf{Hyperparameters} \\ \hline \\[-1.5em]
    Signal-to-noise ratio & $1/\sqrt{2 \eta \kappa\int_0^{\tau_m} \left\vert \beta_e(t) - \beta_g(t) \right\vert^2 \dd t}$ & 1 & $\eta=0.6$, $\kappa / 2\pi = \SI{30}{\mega\hertz}$ \\
    Pulse amplitude (filter) & $\frac{1}{\tau_m} \int_0^{\tau_m} \ReLU(|\Omega_f(t)| - \Omega_\mathrm{max}) \dd t$ & 0.1 & $\Omega_\mathrm{max} / 2\pi = \SI{200}{\mega\hertz}$ \\
    Pulse amplitude (transmon) & $\frac{1}{\tau_m} \int_0^{\tau_m} \ReLU(|\Omega_t(t)| - \Omega_\mathrm{max}) \dd t$ & 0.1 & $\Omega_\mathrm{max} / 2\pi = \SI{200}{\mega\hertz}$ \\
    Forbidden states (transmon) & $\sum_{i=g,e} \sum_{k \geq n}^{N_t-1} \frac{1}{\tau_m} \int_0^{\tau_m} \Tr[\ketbra{k}{k}_t \hat\rho_i(t)] \dd t$ & 1 & $n=2$ or $3$, $N_t=5$ \\
    Forbidden states (undriven normal) & $\sum_{i=g,e} \sum_{k \geq n}^{N_u-1} \frac{1}{\tau_m} \int_0^{\tau_m} \Tr[\ketbra{k}{k}_u \hat\rho_i(t)] \dd t$ & 5000 & $n=2$, $N_u=4$ \\ 
    Critical photon number (resonator) & $\sum_{i=g,e} \frac{1}{\tau_m} \int_0^{\tau_m} \ReLU(\Tr[\hat a^\dagger \hat a \hat\rho_i(t)] - \bar{n}_\mathrm{crit})\, \dd t$ & 0.1 & $\bar{n}_\mathrm{crit} = 16$ \\[0.2em] \hline
    State reset $\ket{g}$ & $\log_{10}(1 - \left\vert \bra{g00} \hat \rho_{g}(\tau_m) \ket{g00}\right\vert^2)$ & 0.3 & -- \\
    State reset $\ket{e}$ & $\log_{10}(1 - \left\vert \bra{g00} \hat \rho_{e}(\tau_m) \ket{g00}\right\vert^2)$ & 1 & -- \\
    State reset $\ket{f}$ & $\log_{10}(1 - \left\vert \bra{g00} \hat \rho_{f}(\tau_m) \ket{g00}\right\vert^2)$ & 0.3 & -- \\ 
    Pulse amplitude (transmon) & $\frac{1}{\tau_m} \int_0^{\tau_m} \ReLU(|\Omega_t(t)| - \Omega_\mathrm{max}) \dd t$ & 0.1 & $\Omega_\mathrm{max} / 2\pi = \SI{600}{\mega\hertz}$ \\[0.2em] \hline\hline
  \end{tabular*}
  \caption{Summary of readout (top) and reset (bottom) cost functions. Each line corresponds to a different contribution to the total cost function, with an overall weight tuned heuristically. All contributions are functions of the problem parameters and/or of the density matrix at certain times. Time integrals are numerically discretized in \SI{1}{\nano\second} time bins. $\ReLU$ denotes a rectified linear unit function such that $\ReLU(x) = 0$ for $x\leq 0$ and $\ReLU(x) = x$ for $x\geq 0$. The states $\ket{k}_t$ and $\ket{k}_u$ denote the $k$-th Fock state in the transmon and undriven normal mode respectively. Finally, $\hat \rho_i$ denotes the density matrix for a transmon initialized in state $\ket{i}_t$.}
  \label{tab:losses}
\end{table*}

\section{Optimal control of a transmon}\label{app:qoc}

\subsection{System parameters}

Unless stated otherwise, we use $E_C / 2\pi = \SI{315}{\mega\hertz}$, $E_J / E_C = 51$, corresponding to a bare transmon frequency $\omega_t / 2\pi = \SI{6}{\giga\hertz}$ and to an anharmonicity $\alpha / 2\pi = \SI{-349}{\mega\hertz}$. The resonator and filter frequencies are $\omega_r / 2\pi = \SI{7.2}{\giga\hertz}$ and $\omega_f / 2\pi = \SI{7.21}{\giga\hertz}$ with couplings $g / 2\pi = \SI{150}{\mega\hertz}$ and $J / 2\pi = \SI{30}{\mega\hertz}$. This yields a transmon-resonator detuning of $\Delta / 2\pi = \SI{1.2}{\giga\hertz}$, a critical photon number of $\bar{n}_\mathrm{crit} = {(\Delta / 2g)}^2 = 16$~\cite{blais2004cavity} and dispersive rates of $\chi / 2\pi = \SI{3.8}{\mega\hertz}$ and $\SI{8.1}{\mega\hertz}$ with the lower and higher normal modes, respectively. Finally, the filter loss rate is $\kappa / 2\pi = \SI{30}{\mega\hertz}$ and the transmon relaxation rate is $\gamma / 2\pi = \SI{8}{\kilo\hertz}$, i.e. $T_1 = \SI{20}{\micro\second}$.

\subsection{Optimization process}

In this work, we optimize transmon readout and reset by simulating the Lindblad master equation of Eq.~(5) with the Hamiltonian of the main text which contains three modes: the full transmon Hamiltonian, a readout resonator and a Purcell filter. In our simulations, we first diagonalize the transmon Hamiltonian in the charge basis using 300 charge states and then truncate the subsystem to $N_t=5$ eigenstates of lowest energy. In parallel, we also diagonalize the resonator-filter subsystem yielding two normal modes -- each coupled to the transmon -- and keep $N_d$ Fock states for the mode used to readout/reset the transmon (driven mode) and $N_u$ Fock states in the other mode (undriven mode). For readout, we typically take $N_d = 85$ and $N_u = 4$. For reset, we instead use $N_d = N_u = 4$ since neither mode is actively driven.

Note that we intentionally select small Hilbert space sizes to lower the simulation runtime: a typical gradient descent on such transmon readout/reset simulations takes from a few hours to a few days per data point using a single GPU~\footnote{GPUs used: Nvidia RTX6000 24GB, Nvidia RTX8000 48GB, Nvidia V100 32GB, Nvidia RTX3080 10GB}. This runtime depends linearly on the number of integration steps per simulation (and thus on the duration of this simulation), on the number of epochs for convergence of the gradient descent, and on the FLOPs of the GPU (when maximally loaded). Importantly, it also depends cubically on the Hilbert space size, for large enough sizes. These dependencies explain the wide discrepency in runtimes between the various simulations performed in the main text.

To avoid probing unphysical simulations with such low Hilbert space sizes, we actively steer the optimizer in two ways. First, we engineer the cost function such as to avoid large populations in boundary Fock states and limit the pulse amplitudes. See~\cref{tab:losses} for detailed cost function contributions. Second, we validate all optimized simulations with a single additional simulation in a larger Hilbert space, using pulses as obtained by the optimization process. For such simulations, we typically take $N_t = 6$, and $N_d = 90$ and $N_u = 8$ for readout, or $N_d = N_u = 8$ for reset. All data points shown in the main text correspond to such validated simulations, and do not deviate significantly from the results obtained on the reduced Hilbert space simulations.

To seed the various optimisations shown in the main text, flat-top envelopes were systematically chosen (for both readout and reset). Regarding drive frequencies, readout drives were seeded at either the bare readout resonator frequency or bare filter frequency. Transmon drives were also seeded at the bare system frequencies, either of the transmon, readout resonator, or filter. The finding of the dressed or stark-shifted drive frequencies can thus be fully attributed to the optimizer. For the transmon reset optimisation, the frequency seeds are chosen through the calibration process discussed in~\cref{app:reset-cal}.

\subsection{Numerical integration scheme}

Regarding the integration scheme of the ordinary differential equation (ODE), we used a second-order Rouchon solver~\cite{le2013low} with a fixed integration time step of $\delta t = \SI{0.003}{\nano\second}$. Such a method ensures preservation of the positivity and trace of the density matrix at all times during the integration, contrary to standard ODE integration schemes. This is particularly interesting in the context of the adjoint state method to avoid divergence during the reverse-time integration of the master equation. Later on in this work, we have also used a standard order-4/5 Dormand-Prince integration scheme~\cite{dormand1980family} which provided better runtimes in the forward pass and similar ones in the backward pass, even considering the stiffness of the reverse-time integration. In any case, several thousands of integration steps are typically required in our simulations. This makes the use of standard automatic differentiation prohibitive in computer memory. A simple pen-and-paper estimation of memory with the previously stated Hilbert space size and $M_t = \num{10000}$ time steps yields a memory footprint of $8 M_t (N_t N_d N_u)^2 = 215\:\mathrm{GB}$ where $8$ is the number of bytes used to store a single-precision complex number.

\subsection{Gaussian filtering}

As stated in the main text, all simulated pulse envelopes are first discretized in $\tau_0 = \SI{1}{\nano\second}$ time bins and then gaussian filtered according to Ref.~\cite{motzoi2011optimal}. Each complex amplitude of the discretized pulse then corresponds to an additional parameter in the optimization. Concretely, the time-dependent pulses read
\begin{equation}\label{eq:filter-pulse}
    \Omega(t) = \mathcal{F}_g\left[ \sum_{j=0}^{\left\lfloor \tau_m/\tau_0 \right\rfloor - 1}  \Omega_j \Pi_j(t, \tau_0) \right] (t)
\end{equation}
where
\begin{equation}
    \Pi_j(t, \tau_0) \equiv \Theta(t - j \tau_0) - \Theta(t - (j+1) \tau_0)
\end{equation}
is the rectangle function, with $\Theta$ is the Heaviside unit step function, and where $\mathcal{F}_g$ is a gaussian filter of bandwidth $\omega_B /2\pi = \SI{250}{\mega\hertz}$. Commuting $\mathcal{F}_g$ with the sum of~\cref{eq:filter-pulse}, we get the simpler expression
\begin{equation}
    \Omega(t) = \sum_{j=0}^{\left\lfloor \tau_m/\tau_0 \right\rfloor - 1} \Omega_j \zeta_j(t),
\end{equation}
where
\begin{equation}
    \zeta_j(t) = \frac{1}{2}\left[\erf\!\left(\omega_0 \frac{t - j\tau_0}{2}\right) - \erf\!\left(\omega_0 \frac{t - (j+1)\tau_0}{2}\right)\right],
\end{equation}
with $\erf$ the error function and $\omega_0 / 2\pi = \SI{425.5}{\mega\hertz}$~\cite{motzoi2011optimal}.

\begin{figure}
   \centering
   \includegraphics[width = \columnwidth]{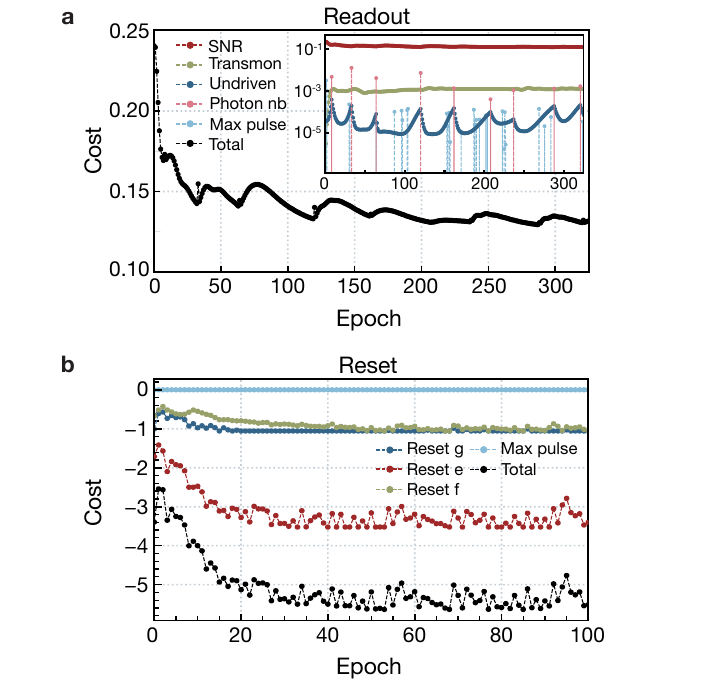}
\vspace{-0.5cm}
\caption{\label{fig:supmat-losses}
   Cost function as a function of the optimization epoch for (a) transmon readout QOC with no transmon drive at \SI{40}{\nano\second}, and (b) transmon reset QOC at \SI{200}{\nano\second}. The inset of panel (a) shows the cost function contributions in log scale, with the main contribution being the inversed SNR. Other contributions are meant to regularize the pulse and to avoid unphysical regions. See~\cref{tab:losses} for the detailed cost function contributions.
}
\end{figure}

\begin{figure*}
    \centering
    \includegraphics[width = \textwidth]{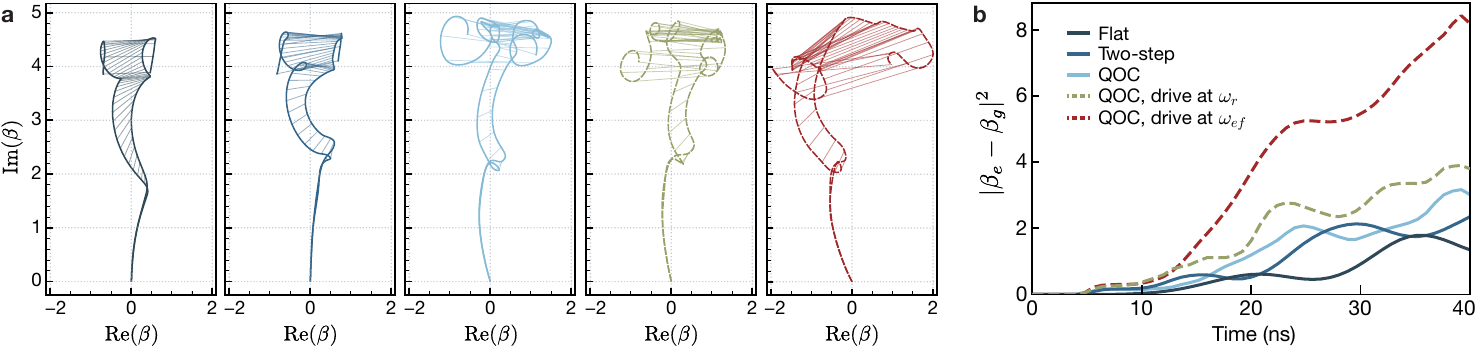}
\vspace{-0.5cm}
\caption{\label{fig:supmat-readout-phasespace}
    (a) Mean phase-space trajectories of the Purcell filter during a \SI{40}{\nano\second} readout. Each panel shows $\beta_e$ and $\beta_g$ against time for a different readout strategy, where $\beta_{e/g} = \operatorname{Tr}[\f \hrho_{g/e}]$ is the filter mean field after initializing the transmon in the $\ket{e}$ or $\ket{g}$ state. Lines that join each trajectory represent \SI{1}{\nano\second} intervals. The optimal control algorithm attempts to maximize SNR, which corresponds to the integrated distance between both trajectories. (b) Phase-space distance against time for the same readout strategies as in (a).
}
\end{figure*}

\subsection{Cost function contributions}

Designing appropriate cost functions is essential to obtain useful quantum optimal control results from numerical optimizations. In this work, we define composite cost functions with a principal component and several smaller regularizing contributions. The cost functions used for transmon readout and reset are summarized in~\cref{tab:losses}. Let us describe each cost function separately.

For transmon readout, the main contribution is the inverse of the SNR. We use the inverse (instead of e.g.~the negation) such that the optimizer focuses on other cost function contributions increasingly more as the SNR grows large. There are five regularizing cost functions on top of this main component. The first two correspond to a maximum pulse amplitude on the filter and transmon respectively, at $\Omega_{max} / 2\pi = \SI{200}{\mega\hertz}$. These cost functions use a rectified linear unit function, or $\ReLU$, such that $\ReLU(x) = 0$ for $x\leq 0$ and $\ReLU(x) = x$ for $x\geq 0$. As such, no hard limits are set on the pulse amplitudes to allow the optimizer to explore a larger parameter space in case interesting solutions can be found at larger amplitudes. Two other contributions correspond to forbidden states in the transmon and undriven normal mode respectively. As previously discussed, these costs are used to avoid probing unphysical pulses that would make use of boundary Fock states sitting close to our artificial numerical truncation. Finally, the last contribution to the cost function is a soft upper bound on the number of photons in the resonator. Indeed, according to the standard theory of dispersive qubit readout~\cite{blais2021circuit}, non-QND readout can be probed by driving the readout resonator above its critical number of photons $\bar{n}_\mathrm{crit}$. This cost function also allows for a fair comparison between all readout strategies by limiting the effective speed of pointer state separation, given by $\bar{n} \chi$ in the absence of additional transmon drives.

For transmon reset, the main contributions are the logarithms of residual population outside the $\ketbra{g00}{g00}$ subspace, for a transmon initialized in either $\ket{g}$, $\ket{e}$, or $\ket{f}$. In particular, we weight these three contributions non-uniformly with a larger weight on the $\ket{e}$ state reset. This is a choice motivated by experimental designs for which population inside the qubit subspace is typically much larger than leaked population at the beginning of resets. However, these weights can be freely adapted to the experiment. The only regularizing cost function for transmon reset is an upper bound on the pulse amplitude, similar to that of transmon readout. Because reset attempts to evacuate the system photons, additional cost functions are not required.

In~\cref{fig:supmat-losses}, we show how typical cost function contributions evolve as a function of the optimization epoch (i.e.~iteration). Panel (a) shows a \SI{40}{\nano\second} transmon readout optimization where only the filter is driven, while panel (b) shows a \SI{200}{\nano\second} reset optimization. In both cases, we find a clear decrease in the total cost in the first 20 to 50 epochs, and convergence to a steady value afterwards. For readout, the cost function is dominated by the SNR cost function. The inset highlights the smaller cost contributions in log scale. For these regularizing cost functions, we regularly see peaks indicating that the optimizer is generating dynamics close to the Hilbert space boundaries before retracting away, all the while achieving increasingly better SNR.

\section{Transmon readout}
\subsection{Assignment error}

In Fig.2(c) of the main text, we show the assignment error of transmon readout as a function of the integration time. This assignment error is computed in Refs.~\cite{gambetta2007protocols,swiadek2023enhancing} assuming that the transmon pointer states stay gaussian during readout. This assumption is verified to a very good approximation for all of our simulated readout sequences. Given this structure, the assignment error reads
\begin{equation}\label{eq:assignerror}
   \varepsilon_a(\tau_m) \simeq \frac{1}{2} \operatorname{erfc}\left(\frac{\mathrm{SNR}(\tau_m)}{2}\right) + \frac{\tau_m}{2T_1},
\end{equation}
where $\varepsilon_a = 1-\mathcal{F} = [P(e|g) + P(g|e)] / 2$, with $P(i|j)$ the probability of measuring state $\ket{i}$ when $\ket{j}$ is prepared, $\mathcal{F}$ the readout fidelity, and $\operatorname{erfc}$ the complementary error function.

\subsection{Phase-space trajectories}

\Cref{fig:supmat-readout-phasespace}(a) shows the readout trajectories obtained by our three optimized control strategies, in comparison with the flat and two-step readout references. Notably, we see that pulse shaping the microwave drives on the cavity and qubit, while still using an experimentally-realistic bandwidth, makes the readout dynamics deviate from the usual dispersive trajectories.
As quantified in~\cref{fig:supmat-readout-phasespace}(b), optimizing the control drives leads to both a faster and a larger separation between the qubit pointer states, which directly translates into better readout SNR.
The red curve, which corresponds to adding the transmon drive at the $\ket{e}$-$\ket{f}$ transition frequency such as to achieve shelving, shows how beneficial that approach is for the system parameters considered here, compared to standard dispersive readout. The discovered scheme of simultaneously shelving and populating the readout mode with photons could straightforwardly be implemented with the typical controls available in current transmon experiments.

\section{Transmon reset}\label{app:reset}
\subsection{Calibration}\label{app:reset-cal}
Since the f0-g1 reset of the transmon $\ket e$ and $\ket f$ states requires two concurrent microwave drives on the transmon, the frequency and amplitude of these drives need to be calibrated in a self-consistent way. In order to get the best reset performance using flat drives on which to compare our optimal controls, we perform a numerical calibration procedure that follows closely what is done in typical experiments~\cite{pechal2014microwave,magnard2018fast}.

\begin{figure*}
   \centering
   \includegraphics[width = \textwidth]{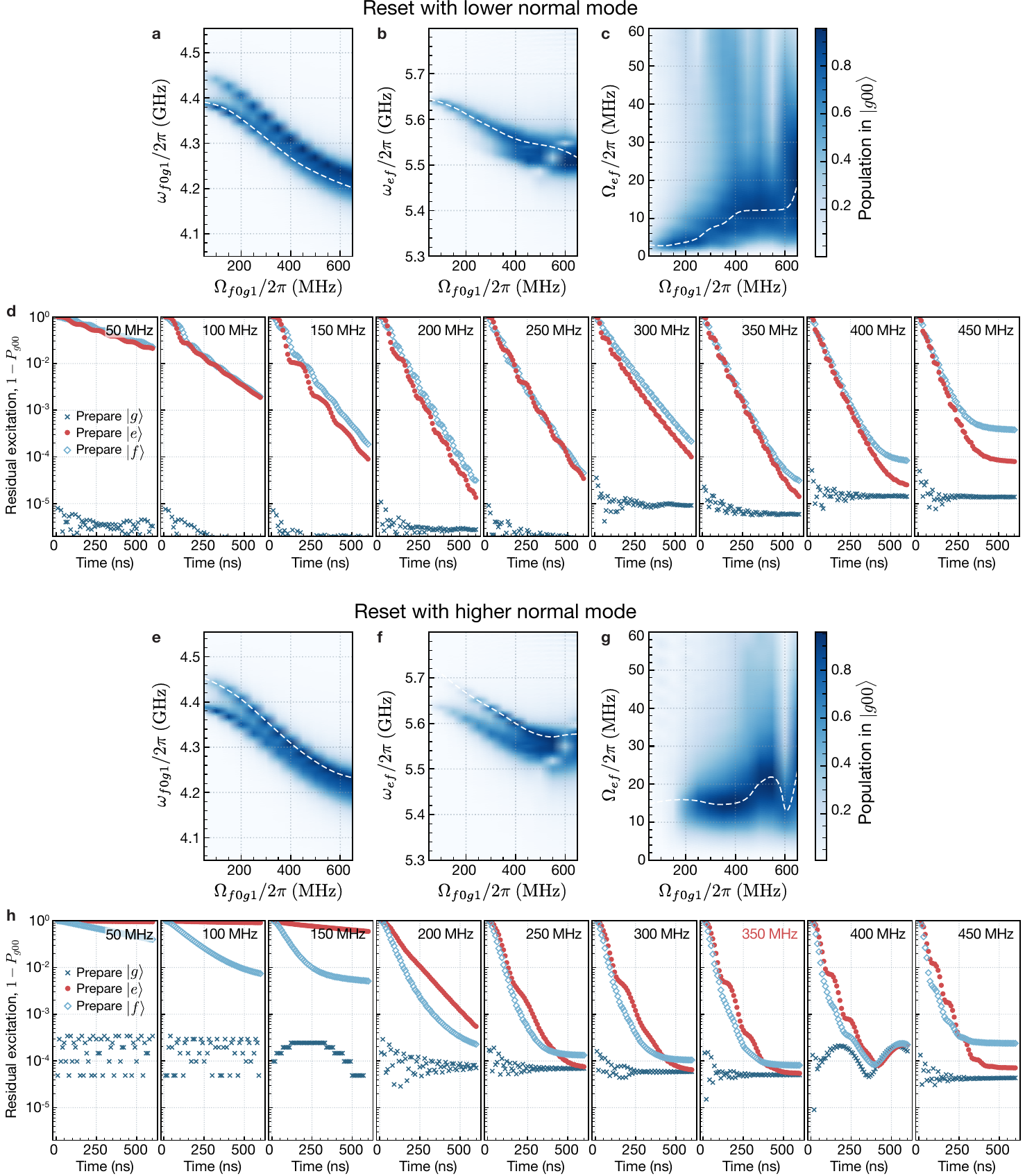}
\vspace{-0.5cm}
\caption{\label{fig:supmat-reset-calibration}
    Numerical calibration of the f0-g1 reset for the reference flat drives.\ (a-c) Following the normal mode of lowest frequency, the f0-g1 drive frequency $\omega_\mathrm{f0g1}/2\pi$, the e-f drive frequency $\omega_\mathrm{ef}/2\pi$, and e-f drive amplitude $\Omega_\mathrm{ef}/2\pi$ are sequentially calibrated.
    See the protocol description in~\cref{app:reset} for more details.\ (d) Resulting f0-g1 reset performance after preparing the $\ket g$, $\ket e$, and $\ket f$ states of the transmon for different f0-g1 drive amplitudes $\Omega_\mathrm{f0g1}/2\pi$ between $\SI{50}{\mega\hertz}$ and $\SI{450}{\mega\hertz}$.\ (e-h) Results for the same calibration protocol, now following the normal readout mode with higher frequency. The selected reference f0-g1 reset for our system is highlighted in red and uses $\Omega_\mathrm{f0g1}/2\pi=\SI{350}{\mega\hertz}$ and the higher normal mode of the coupled resonator-filter subsystem.
}
\end{figure*}

Namely, for a given f0-g1 drive amplitude $\Omega_\mathrm{f0g1}$, we first scan the reset drive frequency $\omega_\mathrm{f0g1}$ to find the ac-Stark shifted resonant frequency that maps the prepared $\ket f$ transmon state back to $\ket g$ after \SI{100}{\nano\second}. The results of this simulated experiment are illustrated in~\cref{fig:supmat-reset-calibration}(a, e) for the lower and higher normal readout mode, respectively, and for different drive amplitudes $\Omega_\mathrm{f0g1}/2\pi$ between $50$ and $\SI{650}{\mega\hertz}$.  Once the frequency $\omega_\mathrm{f0g1}$ is fixed on resonance, we calibrate the $\ket e$-$\ket f$ drive frequency $\omega_\mathrm{ef}$. This is achieved by performing a similar experiment as before, but in which the transmon is prepared in $\ket e$ instead of $\ket{f}$, and both the f0-g1 drive and e-f drives are on. The amplitude ratio $\Omega_\mathrm{f0g1}/\Omega_\mathrm{ef}$ between both drives is kept constant such that the effective f0-g1 Rabi rate yields $\Tilde{g}(\Omega_\mathrm{f0g1})=\Omega_\mathrm{ef}$~\cite{zeytinouglu2015microwave,magnard2018fast}. The results of this second calibration are shown in~\cref{fig:supmat-reset-calibration}(b, f) for the lower and higher modes, respectively. Finally, the $\ket e$-$\ket f$ drive amplitude $\Omega_\mathrm{ef}$ is calibrated from a 1D parameter sweep while performing the same $\ket e$ reset experiment. This sweep is shown in~\cref{fig:supmat-reset-calibration}(c, g). In the end, this whole procedure yields a single set of calibrated parameters $(\omega_{f0g1},\,\omega_{ef},\,\Omega_{ef})$ for every value of $\Omega_{f0g1}$ and for each normal mode. We simulate the corresponding reset of each set of calibrated parameters, and show the corresponding results in the panels of~\cref{fig:supmat-reset-calibration}(d, h).

Out of all of the calibrated sets of parameters, we selected the best performing reset (both in terms of speed and fidelity) to compare with our optimal controls in Fig.~3 of the main text. This corresponds to a f0-g1 amplitude of $\Omega_\mathrm{f0g1}/2\pi = \SI{350}{\mega\hertz}$ driving the higher frequency normal mode of the coupled resonator-filter subsystem. This mode has a larger dispersive coupling to the transmon than the lower frequency normal mode. This particular reset is shown with a red title in~\cref{fig:supmat-reset-calibration}(h).

\subsection{Pulses}
\begin{figure*}
   \centering
   \includegraphics[width = \textwidth]{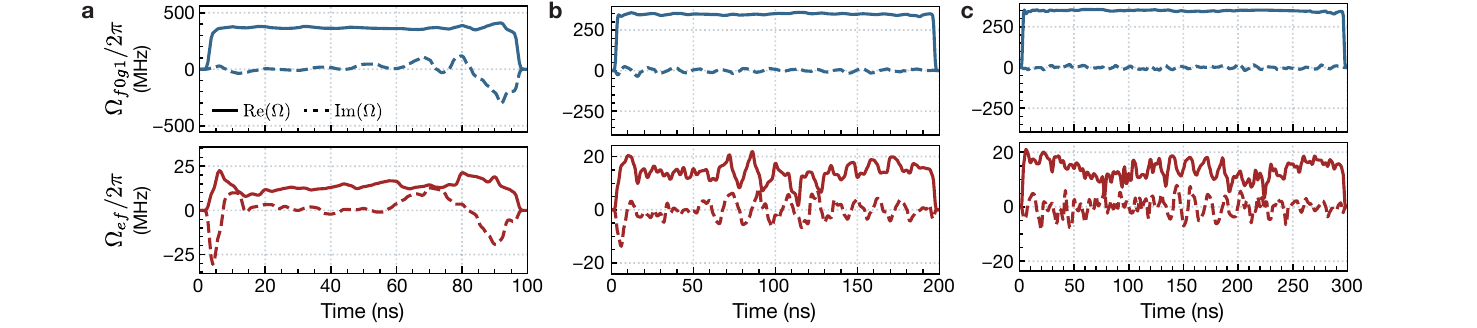}
\vspace{-0.5cm}
\caption{\label{fig:supmat-reset-pulses}
   Optimal pulse shapes for reset durations of (a) $\SI{100}{\nano\second}$, (b) $\SI{200}{\nano\second}$, and (c) $\SI{300}{\nano\second}$. The f0-g1 drive $\Omega_\mathrm{f0g1}$ is illustrated in blue and the e-f drive $\Omega_\mathrm{ef}$ in red. Recall that although the features of $\Omega_\mathrm{ef}$ appear to be sharp, the pulses are filtered using an experimentally conservative gaussian filter with a bandwidth of $\SI{250}{\mega\hertz}$, here and in all of our simulations. This realistic filtering makes these pulses easily implementable in practice.
}
\end{figure*}

In~\cref{fig:supmat-reset-pulses}, we present the optimal reset pulse shapes found using our quantum optimal control approach. The \SI{200}{\nano\second} pulse of panel (b) produces the transmon population dynamics illustrated with the pink markers in Fig.~3 of the main text. Interestingly, in the \SI{100}{\nano\second} pulse of panel (a), we see that the numerical optimization yields an initial $\pi$-pulse to transfer population from $\ket e$ to $\ket f$ which quickly decays back to $\ket g$ because of the f0-g1 drive. This behaviour can be understood by the fact that our chosen cost function puts more weight into resetting the $\ket e$ state. As such, the optimal drive scheme with this constraint initially resets the $\ket e$ state, and subsequently resets the population that was initially in the $\ket f$ state at the beginning of the protocol. This behaviour is consistent with the fact that the residual population when initializing in $\ket e$ converges after \SI{100}{\nano\second} whereas it converges after \SI{200}{\nano\second} for the $\ket f$ initial state, as reported in the main text. Our choice of relative weights between the different residual populations is motivated by the thermal population of a high frequency transmon (\SI{5}{\giga\hertz}) in a relatively cold environment ($\SI{20}{\milli\kelvin}\sim\SI{417}{\mega\hertz}$), but is arbitrary and could easily be adapted to other experimentally relevant scenarios, e.g.~a low-frequency fluxonium qubit (<\SI{1}{\giga\hertz}) in contact with the same thermal bath.

\end{document}